\documentclass[manuscript]{emulateapj}

\shorttitle{IC 10 Star Clusters}
\shortauthors{Lim \& Lee}

\begin{document}


\title{
The Star Cluster System in the Local Group Starburst Galaxy IC 10
}


\author{Sungsoon Lim\altaffilmark{1,2,3} and
Myung Gyoon Lee\altaffilmark{1}} 

\email{slim@pku.edu.cn and mglee@astro.snu.ac.kr}
\affil{$^1$Astronomy Program, Department of Physics and Astronomy, Seoul National University, Seoul 151-747, Korea \\
$^2$Department of Astronomy, Peking University, Beijing 100871, China \\
$^3$Kavli Institute for Astronomy and Astrophysics, Peking University, Beijing 100871, China}

%
%




\begin{abstract}
We present a survey of star clusters in the halo of IC 10, a starburst galaxy in the Local Group based on Subaru $R$ band images and NOAO Local Group Survey $UBVRI$ images. 
We find five new star clusters. 
All these star clusters are located far from the center of IC 10, 
while previously known star clusters are mostly in the main body.
Interestingly the distribution of these star clusters shows an asymmetrical structure elongated along the east and south-west direction.
We derive $UBVRI$ photometry of 66 star clusters including these new star clusters as well as previously known star clusters. 
Ages of the star clusters are estimated from the comparison of their $UBVRI$ spectral energy distribution with the simple stellar population models. 
We find that the star clusters in the halo are all older than 1 Gyr, while those in the main body have various ages from very young (several Myr) to old ($>1$ Gyr).
The young clusters ($<10$ Myr) are mostly located in the H$\alpha$ emission regions and are concentrated on a small region 
at $2\arcmin$ in the south-east direction from the galaxy center, 
while the old clusters 
are distributed in a wider area than the disk.
Intermediate-age clusters ($\sim100$ Myr) are found in two groups. One is close to the location of the young clusters and the other is at $\sim 4\arcmin$ from the location of the young clusters. The latter may be related with past merger or tidal interaction.

\end{abstract}





\section{Introduction}
The starburst event is important to understand evolution of galaxies.
Most galaxies experience starburst phases in their evolutionary histories, and present-day stars might have been formed mostly in the starbursts (e.g. \citealp{Elb03}).
Massive stars formed from the starbursts play an important role in stellar feedback such as chemical enrichment. 
Therefore, the starburst is one of the main events for evolution of galaxies, and a study of nearby starburst galaxies can provide insights into the processes of the starburst.

One of the robust definitions for starburst galaxies is that galaxies have much higher star formation rate than that of their past average (e.g. \citealp{Meu95}). We need to derive star formation history and current star formation rate of galaxies to determine the starburst galaxies by this definition (e.g. \citealp{McQ10}). However, it is hard to derive star formation histories of galaxies, so the starburst galaxies can be categorised by finding other properties of the starburst events such as high star formation rates, existence of massive stars, and presence of super star clusters.

IC 10 is the nearest starburst galaxy \citep{Mas02,Kim09}.
An unusually large number of HII regions \citep{Hod90}, and far-IR luminosity \citep{Mel94} in this galaxy show that it has a much higher star formation rate than other dwarf galaxies in the Local Group.
IC 10 has a large number of Wolf-Rayet (W-R) stars, considering its total luminosity \citep{Mas02}.  
The surface number density of W-R stars in IC 10 is about one hundred times higher than that of the Large Magellanic Cloud, and it is the highest among the Local Group galaxies, showing that the star formation in IC 10 is on the burst mode \citep{Mas95,Mas02}.

IC 10 is also known as the nearest blue compact dwarf galaxy (BCD) based on its surface brightness and structure. The $B-$band surface brightness of IC 10 is much higher than that of other dwarf galaxies, and it is similar to that of BCDs \citep{Ric01}. BCDs possess underlying old stellar populations with low surface brightness (e.g. \citealp{Dro01,Cro02}). IC 10 also has old stellar populations such as red giant branch (RGB) stars and C stars \citep{Dem04,Kim09,San10}, and they are much more extended and much fainter than the young stellar main body. BCDs mostly have very high rates of star formation as suggested by their strong H$\alpha$ emission \citep{Gil03}, and this property of BCDs is similar to that of the starburst galaxies. These properties of BCDs are also similar to those of IC 10.

To study formation and evolution of IC 10, several studies focused on the star clusters in IC 10.
\citet{Kar93} and \citet{Geo96} presented the first list of seven star cluster candidates in IC 10 detected in the ground-based images, including only their positions. 
\citet{Hun01} found 13 stellar associations and clusters from $F336W$, $F555W$, $F814W$, and $F656N$ images of the eastern part of the main body of IC 10 taken with Wide Field Planetary Camera 2 (WFPC2) on the Hubble Space Telescope (HST). Two of them are old (with ages$>350$Myr) and the rest  are young (ages with 4--30 Myr). They have half-light radii of
1.5--6.0 pc and absolute magnitudes of $-10.0<M_V<-6.6$. 
\citet{Hun01} suggested that these young star clusters might have been formed during the recent starburst phase.
Later \citet{Tik09} searched for star clusters in IC 10, using WFPC2 and Advanced Camera for Survey (ACS)/Wide Field Channel (WFC) images of the main body of IC 10, and WFPC2 images of two outer fields  in the HST archive, finding 57 star clusters.
They confirmed that four of the seven star cluster candidates  in \citet{Kar93} covered by the HST images are indeed star clusters.
They found that 
two of the star clusters in \citet{Hun01} are single stars, and two other clusters in \citet{Hun01} are part of an extended star complex.
They classified roughly the star clusters into three age groups based on their morphology: young (34), intermediate-age (5), and old (18).
However, no photometry data is available for these star clusters.
More recently, \citet{Sha10} provided a catalog of star clusters in IC 10 including three band photometric data based on the same HST images as used in \citet{Tik09}.  
 
These previous studies covered mainly the main disk region of IC 10 so that little is known about the star clusters in the outer region of this galaxy. 
The existence of the RGB stars in the outer region \citep{Dem04,San10} suggests that there may be significant population of old star clusters in the outer region.
In addition, photometric information is not enough to study physical properties for most of the previously known star clusters. 
In this study we survey the outer region of IC 10 to find any star clusters in the halo using the wide field images available in the archive.
We also derive homogeneous photometric magnitudes of all star clusters in IC 10 and investigate their photometric properties.

This paper is organized of as follows.
Section 2 describes data, 
the method of star cluster selection, 
how to derive the integrated photometry
of the star clusters, and 
how to estimate ages of the star clusters.
\S 3 presents  a catalog of star clusters newly discovered as well as previously known in IC 10. Then we show color-magnitude diagrams, color-color diagrams, age distribution, and spatial distribution of the star clusters.
Implications of the results are discussed in \S 4,
and primary results are summarized in the final section.

We adopt a distance to IC 10, 715 kpc ($(m-M)_0=24.27\pm0.03$) and a foreground extinction values, $E(B-V)=0.52\pm0.04$ that were determined by the measurement of $K_S$-band luminosity of the tip of the red giant branch, and $UBV$ photometry of early-type stars, respectively \citep{Kim09}. 
At this distance, one arcsecond corresponds to 3.5 pc, and one arcminute is 208 pc.
$D_{25}$ of IC 10 is $6\arcmin.3$ \citep{deV91} that is 1.3 kpc at the distance of IC 10. 
The star formation rate of IC 10 is $0.6 M_{\odot}/yr$, derived from Br$\gamma$ IR imaging \citep{Bor00}, and the surface star formation rate of IC 10 is $0.03 M_{\odot} yr^{-1} kpc^{-2}$, based on the integrated H$\alpha$ luminosity of IC 10 \citep{Hun01}.
However, we adopted the surface star formation rate, $0.08 M_{\odot}yr^{-1}kpc^{-2}$ that is a logarithmic mean value of those in \citet{Bor00} and \citet{Hun01}.

\section{Data Reduction \& Analysis}

\subsection{Data}

We use $UBVRI$ images of IC 10 in the Local Group Survey (LGS, \citealp{Mas07}) and $R$ band Suprime-cam images of IC 10 in the Subaru archive (SMOKA).
The LGS images cover a $37\arcmin \times 37\arcmin$ field with a pixel scale of
$0\arcsec.27$/pixel, and the Subaru images have $29\arcmin \times 37\arcmin$ field of view 
with a pixel scale of $0\arcsec.20$/pixel. 
The seeing values of the images are about $0\arcsec.9$ and $0\arcsec.7$ for the LGS and Suprime-cam images, respectively.
The field of view of the overlapped region between these two fields is about $25\arcmin \times 34\arcmin$,
 and we searched for  star clusters in this overlapped region.
The positions of these fields are marked in Figure 1.

\begin{figure}
 \epsscale{1.0} \plotone{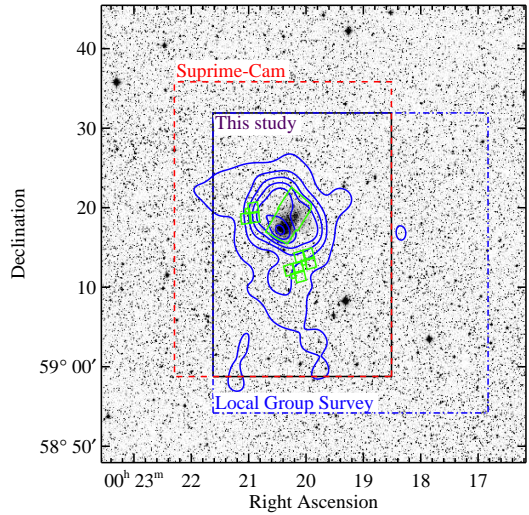}
 \caption{A grey scale map of the digitized sky survey image of IC 10 showing the positions of the fields used in this study: Subaru field (dashed line), NOAO Local Group Survey field (dot-dashed line), and HST fields (small polygons). Solid line contours represent an HI map from \citet{van02}. 
 The solid line box represents the survey area in this study
\label{finder}}
\end{figure}


\subsection{Star Cluster Selection}
Typical star clusters appear to be slightly extended in the Subaru R-band images, but not in the LGS images. In addition, a small number of individual faint stars in the outskirts of star clusters in IC 10 are resolved in the Subaru images because of its depth and better seeing. Therefore we used Subaru $R$-band images for star cluster selection. However, Subaru images have only in $V$,$R$, and $I$ band data, so they are limited in estimating star cluster ages. We used the LGS images to extend the wavelength coverage of the SEDs for star clusters. Therefore we selected star clusters in IC 10 with following steps.

\begin{figure}
 \epsscale{1.0} \plotone{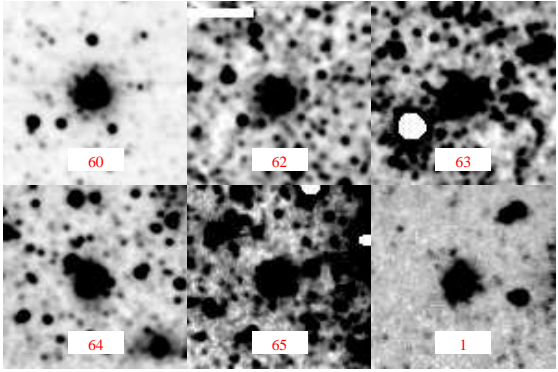}
 \caption{Gray scale maps of thumbnail $R$ band Suprime-cam images of five newly found star clusters in IC 10, and one known star cluster (ID 1). Field of view of each image is 10\arcsec by 10\arcsec. North is up and east to the left.
\label{thu}}
\end{figure}

We ran Source Extractor \citep{Ber96} to the LGS $R$-band images of IC 10 and made a list of extended sources with small stellarity values ($<0.8$).
Then we inspected the images of the bright extended sources with $R<21.0$ mag (similar to the magnitude limit of the known star clusters in IC 10) in the
Subaru $R$-band images to select star clusters according to the selection criteria: (a) circular shape of the sources, and (b) the existence of some resolved stars in the outer region of each source. 
We searched only the outer region of this galaxy, because the star clusters in the main body of the galaxy were already found in previous studies based on HST images \citep{Hun01,Tik09}.
Finally, we selected seven star clusters. Two of these were found in \cite{Kar93} and the rest are new ones. 
Thumbnail images of the five new star clusters are shown in Figure \,\ref{thu}.
We also show a thumbnail image of one known star cluster (No. 1 in \citet{Tik09}) for comparison. 

\subsection{Photometry and Size Estimation}

We derived instrumental $UBVRI$ magnitudes of the five new star clusters as well as the 61 known star clusters \citep{Kar93,Tik09} using an aperture with $1\arcsec.6$ radius from the LGS  images with PHOT/DAOPHOT task in IRAF\footnote{IRAF is distributed by the National Optical Astronomy Observatories, which are operated by the Association of Universities for Research in Astronomy, Inc., under cooperative agreement with the National Science Foundation.}. 
We used annuli of $5\arcsec.4$--$8\arcsec.1$ for sky estimation.
Then we transformed these magnitudes on the standard system using the instrumental magnitudes of bright point sources in the same images and the $UBVRI$ photometry catalog of the point sources in IC 10 given by \citet{Mas07}. 
Full Width at Half Maximum (FWHM) values of $UBVRI$ band images are similar ($0\arcsec.9$-$1\arcsec.0$), and we used $1\arcsec.6$ radius aperture to derive the color of star clusters. Therefore the missing flux due to the use of a fixed aperture photometry is negligible.

The star clusters in IC 10 appear as extended sources in the images so that we need to apply different aperture correction depending on the cluster size. 
We adopted the aperture correction method that was applied by \citet{Hwa08} and \citet{Lim13}. 
This method uses the correlation between the slope of light profiles and the calculated aperture correction values. 
We calculated the slope of light profiles using magnitudes of 6 pixel and 10 pixel apertures.
The aperture correction values were calculated from the magnitude difference of 6 pixel and 20 pixel apertures derived for the isolated and bright star clusters.
We restrict the maximum correction values as 0.6 mag to avoid over correction.
The aperture correction can be applied to all $UBVRI$ magnitudes, but the aperture correction for each filter may increase color errors due to their uncertainties. In addition, the FWHM values for $UBVRI$ band images are similar ($0\arcsec.9$-$1\arcsec.0$). Therefore, we apply the aperture correction only for $V$ to derive the total $V$-band magnitudes, and using the same colors for SEDs.

We compared our photometry with that of \citet{Sha10}. There are differences in both $V$-band magnitudes and ($V-I$) colors between the two studies. The mean value of the difference in the $V$-band magnitudes (This study minus \citealp{Sha10}) is -0.25 with the standard deviation of 0.33, and the mean value of the difference in the ($V-I$) colors is 0.25 with the standard deviation of 0.17. These differences may be mainly due to different filter systems or aperture sizes. We used Johnson’s $UBVRI$ filter system, while they used the HST filter system. In addition, they used both $F555W$ and $F606W$ filters for $V$-band, but they did not distinguish $F555W$ and $F606W$ filters in their catalog. \citet{Sha10} did not provide any information on the aperture sizes so that we cannot discuss it more.

We estimated the sizes of the star clusters using 2-dimensional fitting tool ISHAPE \citep{Lar99} on the Suprime-cam $R$-band images. 
The MOFFAT15 function is adopted to estimate the FWHM of the star clusters.
We derived effective radii of the star clusters using $r_{\rm eff} = 1.13 {\rm FWHM}$.
Several clusters were saturated on this image so that we could not obtain their sizes.
Finally, we derived the sizes of 58 star clusters among the cluster sample. 

\subsection{Estimation of Star Cluster Age }

We derived ages, masses, and extinction values of the star clusters by comparing $UBVRI$ spectral energy distribution (SED) of the star clusters with simple stellar population (SSP) models given by \citep{Bru03}.
We assume a Salpeter initial stellar mass function (IMF) of $N(m)dm \varpropto m^{-2.35}$ with the mass range from $0.1 M_\odot$ to $100 M_\odot$.
The Salpeter IMF was often adopted in the previous studies (see \citealp{Lim13}) so that we adopt it  for comparison with the previous studies. 
The effect of assuming different IMFs is little for the optical colors of SSP models \citep{Lim13}, so that age distributions of the star clusters are not much affected by the different IMFs.
The metallicity of IC 10 is known to be similar to that of SMC ($Z=0.004$, \citet{Ski89,Gar90,Mag03}), but \citet{Hun01} suggested that the integrated $UVI$ colors of the star clusters in IC 10 follow the $Z=0.008$ evolutionary track better than the $Z=0.004$ track in the color-color diagrams.
Therefore, we assumed two values for the metallicity, $Z=0.004$ and $Z=0.008$ for SED fitting.
\citet{Kim09} suggested that the foreground reddening value of IC 10 is $E(B-V)=0.52\pm0.04$ and the total reddening value including the internal reddening of IC 10 is $E(B-V)=0.98\pm0.06$.
Therefore, we adopted the foreground reddening value, $E(B-V)=0.52$, and assigned  the range of internal reddening value $E(B-V)=0.0$ to 0.6.

\section{Results}

\subsection{A Catalog of the Star Clusters in IC 10}

We made a catalog of 66 star clusters in IC 10 including 61 known star clusters \citep{Kar93,Hun01,Tik09} and 5 new star clusters found in this study, as listed in Table 1.
Table 1 includes positions, $UBVRI$ photometry, ages and reddening values of the star clusters.
We investigated optical properties of these star clusters using this catalog.

\begin{figure}
 \epsscale{1.0} \plotone{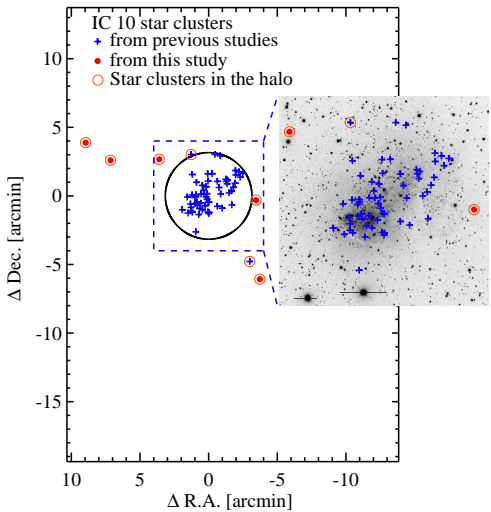}
 \caption{Spatial distribution of the star clusters in IC 10. Crosses and filled circles show the known star clusters and the new star clusters, respectively. A large circle shows a main body region with radius of $R_{25}=3\arcmin.15$. Small circles represent the star clusters in the halo. The box with the grey scale image shows zoom-in of the main body of IC 10. The grey scale image is the LGS $R-$band image.
\label{spa}}
\end{figure}

Figure \ref{spa} shows a spatial distribution of the star clusters in IC 10. 
We divided the star cluster sample into two groups according to their position relative to R$_{25}$ of IC 10 ($R_{25} =3\arcmin.15$, $\approx 655$ pc): 
the main body star clusters at $R<3\arcmin.15$
and the halo star clusters at $R \geq 3\arcmin.15$. 
Several features are noted in this figure.
First, most star clusters are distributed in the main body, forming an elongated structure from south-east to north-west. 
Second, four of the halo star clusters are located along the east direction,
and two are located at $6\arcmin$-$7\arcmin$ south east direction from the galaxy center.
Third, the most distant star cluster (ID No. 64) is at 9.8 arcmin ($\sim2$ kpc) from the galaxy center.

\begin{figure}
 \epsscale{1.0} \plotone{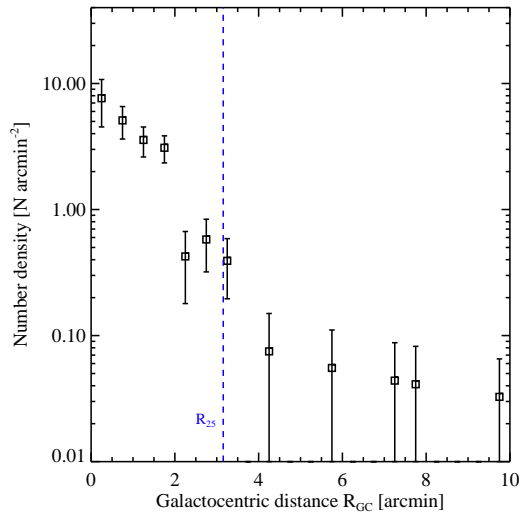}
 \caption{Number density profiles of the star clusters in IC 10. 
Open squares and error bars represent number density and its error of star clusters, respectively. 
The vertical dashed line shows the location of R$_{25}$ for IC 10. 
\label{np}}
\end{figure}

Number density profiles of the star clusters in IC 10 are shown in Figure \ref{np}. 
The number density profile decreases slightly with the galactocentric distance
at $R=2\arcmin$, and drops but stays almost constant at $2\arcmin < R <4\arcmin$. 

\subsection{Color-Magnitude Diagrams, Color-Color Diagrams and Sizes of the Star Clusters}

\begin{figure}
 \epsscale{1.0} \plotone{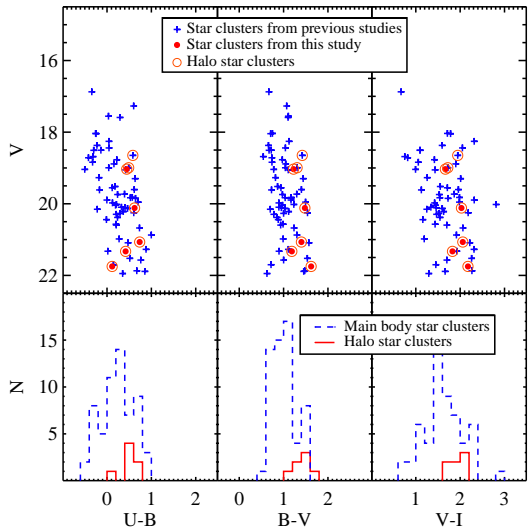}
 \caption{Color-magnitude diagrams of the star clusters in IC 10 (upper panel), and color histograms of the bright star clusters with $V<21$ mag (lower panel). Dashed and solid lines represent histograms of the star clusters in the main body and halo, respectively. Crosses and filled circles represent known star clusters and newly found star clusters, respectively. Open circles represent the halo star clusters. \label{cmd}}
\end{figure}

Figure \ref{cmd} displays color-magnitude diagrams (CMDs) and color histograms of the star clusters in IC 10.
The brightest star cluster in the main body is as bright as $V\sim17$ ($M_V\sim-9$ after correction of the foreground reddening), while that in the halo is as bright as $V\sim18.5$ ($M_V\sim-7.4$, after correction of the foreground reddening).
The color ranges of the star clusters are $-0.5\lesssim(U-B)\lesssim1.0$, $0.5\lesssim(B-V)\lesssim2.0$, and $0.5\lesssim(V-I)\lesssim3.0$.
The ($B-V$) color histogram of the main body star clusters shows two peaks at ($B-V$)$=1.1$ and ($B-V$)$=1.5$, while that of the halo star clusters has only one peak at ($B-V$)$=1.5$.
The star clusters in the halo are mostly redder and fainter than those in the main body.
Above result suggests that the halo star clusters are mostly older than those in the main body. However the reddening effect should be considered.
If we adopt the foreground extinction with $E(B-V)=0.52$ \citep{Kim09}, the reddening-corrected ($B-V$) colors of the halo star clusters are mostly $(B-V)_0 \leq 0.9$, which is similar to that of the metal-poor globular clusters in the Milky Way.

\begin{figure}
 \epsscale{1.0} \plotone{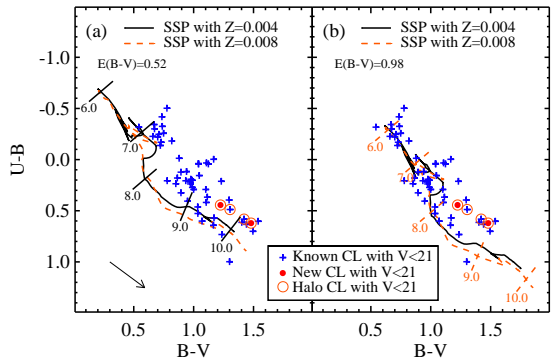}
 \caption{($U-B$)-($B-V$) color-color diagrams of the bright star clusters with $V<21.0$ mag in IC 10.
Symbols are the same as in Figure \ref{cmd}. 
Solid lines and dashed lines represent the simple stellar population (SSP) models for Z=0.004 and 0.008 \citep{Bru03}, respectively. They are shifted according to the reddening of $E(B-V)=0.52$ in (a), and $E(B-V)=0.98$ in (b). Arrows represent the reddening direction. Numbers along the SSP models represent log(age(yr)).
\label{ubvccd}}
\end{figure}

Figure \ref{ubvccd} shows ($U-B$)-($B-V$) color-color diagrams of the bright star clusters ($V<21$) in IC 10.
We compared the color-color diagrams  with SSP models \citep{Bru03}. 
Distinguishable features are as follows.
First, the color-color diagram of the main body star clusters follows the SSP model when the model was shifted with the total extinction value 
($E(B-V)=0.98$), while the color-color diagram of the halo star clusters matches the SSP model when only the foreground extinction was corrected ($E(B-V)=0.52$).
Second, the star clusters in the main body follows the SSP model from young ages (1--10 Myr) to intermediate ages ($\sim1$ Gyr) when the foreground extinction and internal extinction are adjusted on the SSP model.
Third, the halo star clusters are concentrated on the old age area of the SSP model with the foreground extinction correction.
The color-color diagram  of the star clusters gives a hint for their ages and extinctions. However, there is age-extinction degeneracy,  which makes hard to derive ages and extinctions of the star clusters separately.  
Therefore, we study ages of the star clusters quantitatively using the SED fit method in the next section.

\begin{figure}
 \epsscale{1.0} \plotone{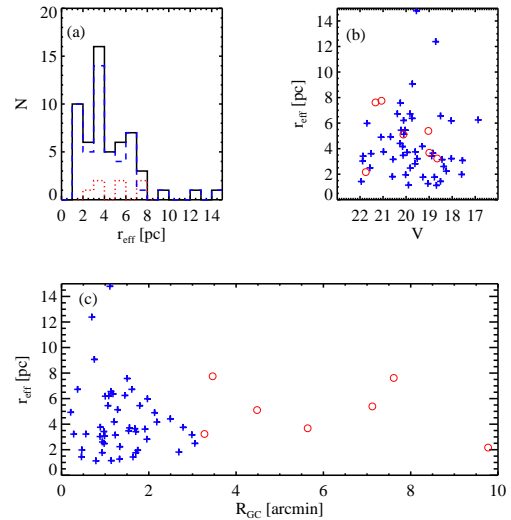}
 \caption{Sizes (effective radii) of the star clusters in IC 10. (a) Size distribution of the star clusters in IC 10. Solid, dashed, and dotted lines represent all, main body, and halo star clusters. (b) Size versus $V$ band magnitude of the star clusters. Crosses and open circles represent main body and halo star cluster, respectively. (c) Size of the star clusters versus galactocentric distance.
\label{siz}}
\end{figure}

Figure \,\ref{siz}(a) shows size distributions of the star clusters.
Effective radii of the star clusters are mostly ranging from $0.5$ pc to 8 pc, with a few larger than 8 pc. 
The peak value of the size distribution is about 3.5 pc, which is similar to those of the star clusters in M51 and M82 \citep{Hwa08, Lim13}.
The sizes of the star clusters in the halo are similar to those of the star clusters in the main body.
There is little correlation between the magnitudes and sizes of the star clusters (Figure \,\ref{siz}(b)).
We also investigate the relation between the sizes and galactocentric distance (Figure \,\ref{siz}(c)), finding that the size of the star clusters changes little depending on the galactocentric distance. 

\subsection{Ages of the Star Clusters}

\begin{figure}
 \epsscale{1.0} \plottwo{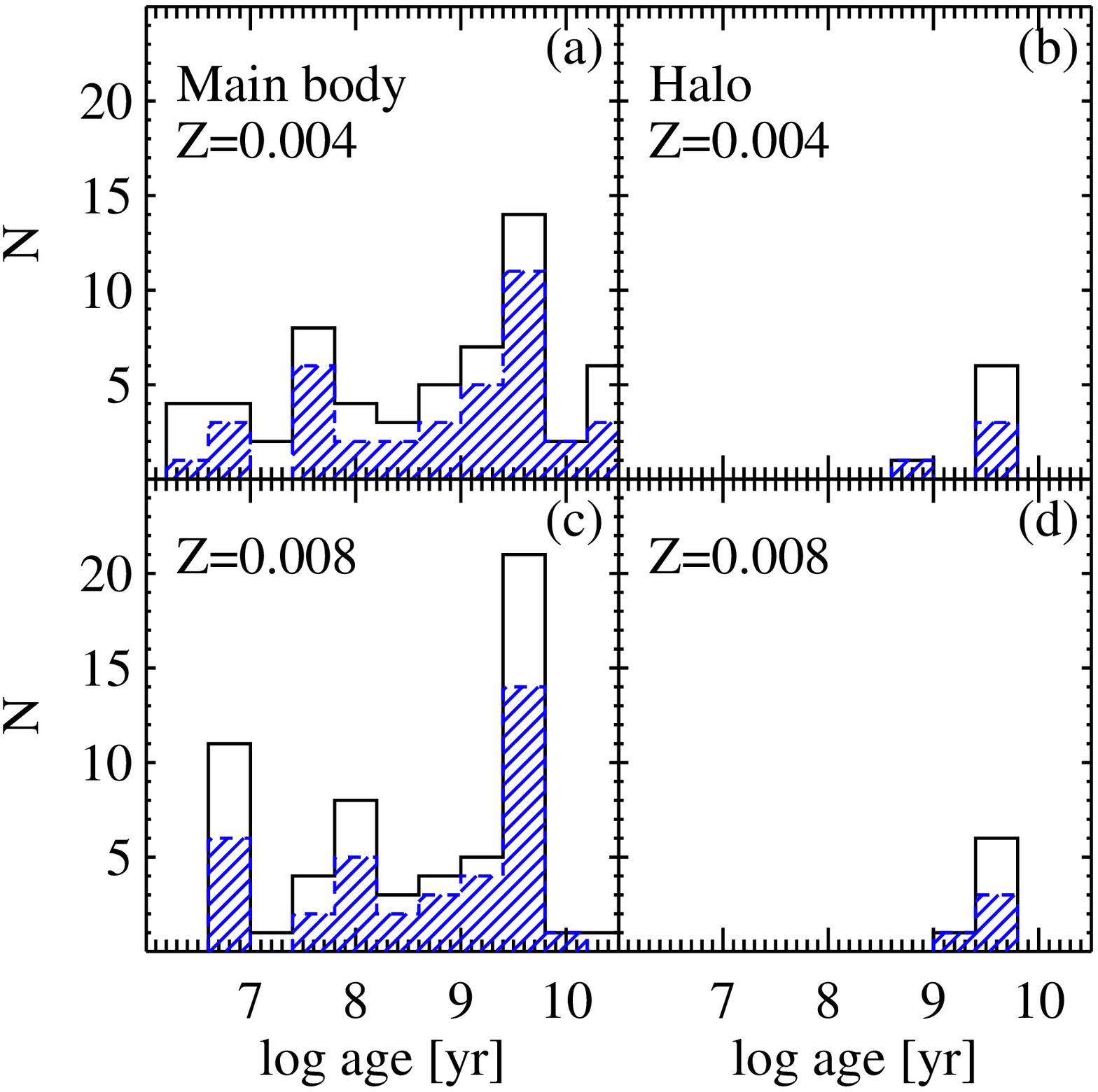}{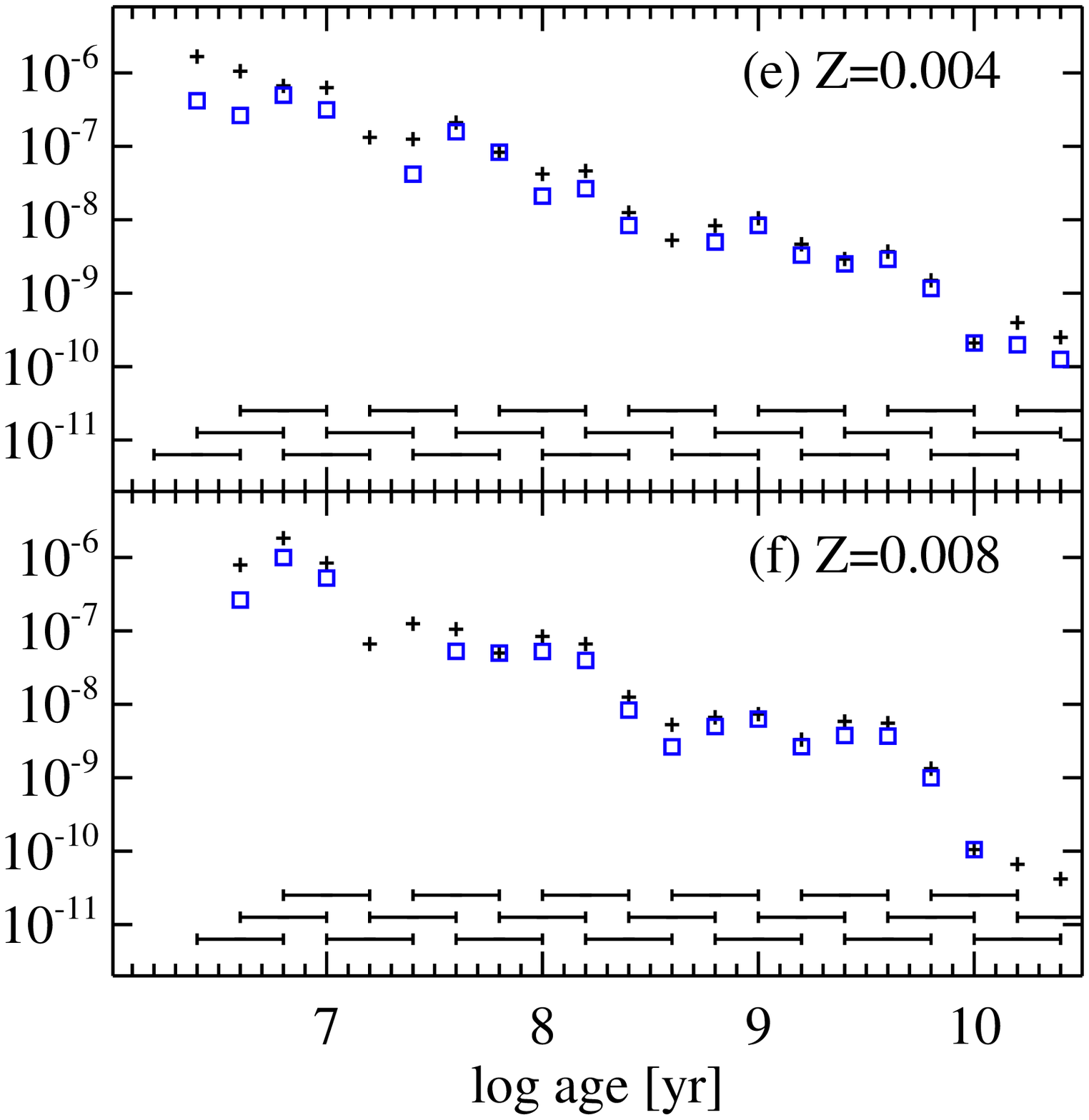}
 \caption{$left$: Age distributions of the main body (a,c) and halo (b,d) star clusters in IC 10 derived for Z$=0.004$ (a,b) and Z$=0.008$ (c,d), respectively. Solid line histograms and hatched histograms  represent the age distributions of all star clusters and those with good fits, respectively. $right$: Age distributions of the main body star clusters with parameters of dN/dt derived for Z$=0.004$ (e) and Z$=0.008$ (f), respectively. Pluses and open squares represent all ages of star clusters and well fitted  ages of star clusters. Error bars at the bottom of figures show the sizes of the bins. 
\label{age1}}
\end{figure}

The left panels in Figure \ref{age1} display age distributions of the star clusters in the main body and halo region. It shows two distinguishable features. First, the ages of the star clusters in the main body span a large range of ages from young ($\sim5$ Myr) to old ($>1$ Gyr), while the ages of the halo star clusters are mostly old ($>1$ Gyr). Second, the age distributions appear to be varying, depending on the metallicities, but their peak positions at 6 Myr, 100 Myr, and 4 Gyr are consistent. The youngest peak at 6 Myr is consistent with the epoch of the recent starburst in IC 10 \citep{Hun01,Vac07}. It is noted that this normal age distribution may have effects due to sample size and logarithmic binning. Therefore, we plotted the age distribution with the number of star clusters per unit time in the right panels of Figure \ref{age1}. It is noted that the two young age peaks are shown in both age distributions with logarithmic binning and normalized numbers. However, they also suffer from stochastic effects of low mass star clusters. We divide the star clusters into three groups according to their ages: young group ($\leq 10$ Myr), intermediate-age group ($10$ Myr --$1$ Gyr), and old group ($> 1$ Gyr).

\begin{figure}
 \epsscale{1.0} \plotone{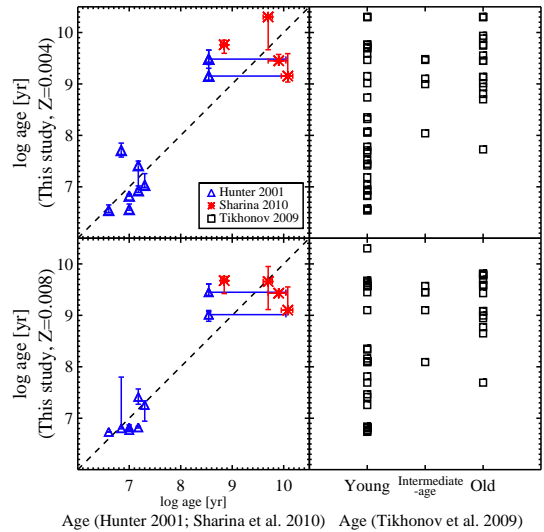}
 \caption{Comparison of the star cluster ages with previous studies. The left column  shows ages from this study versus ages from \citet{Hun01} and \citet{Sha10}, and the right column displays the ages from this study versus the ages from \citet{Tik09}. The top row and bottom row show the ages of the star clusters in this study derived for Z$=0.004$ and Z$=0.008$, respectively.
\label{comp1}}
\end{figure}

We compared the ages of the star clusters with previous studies \citep{Hun01,Tik09}.
\citet{Hun01} obtained the ages of seven star clusters in the cluster sample, using the $(U-V)$--$(V-I)$ color-color diagram compared with theoretical SSP models from \citet{Lei99} for Z$=0.004$ and Z$=0.008$.
She derived ages of two resolved star clusters using CMDs of the resolved stars in each cluster. We compare the ages of these nine star clusters with our results in Figure \,\ref{comp1}(a).
The ages derived in this study are consistent with those of \citet{Hun01}.
\citet{Tik09} distinguished ages of the 56 star clusters by their morphology, and they categorized their cluster sample as young, intermediate-age, and old star clusters. We compare the ages of these star clusters with our results in Figure \,\ref{comp1}(b).
It shows that there is a weak correlation between the two results.
\citet{Sha10} provided the age information for four star clusters in their sample, and they are mostly old ($> \sim1$ Gyr). We compared the ages of these star clusters with those in this study, as plotted in Figure \,\ref{comp1}(a).  Their age estimates for these star clusters are consistent with those in this study.

\begin{figure}
 \epsscale{1.0} \plotone{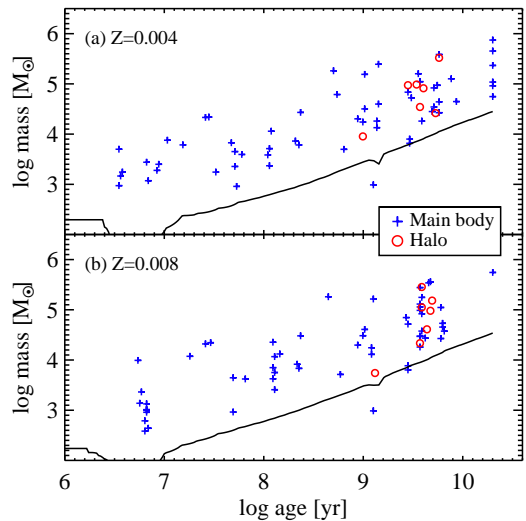}
 \caption{Mass and age relations of the main body (pluses) and halo (circles) star clusters derived for Z$=0.004$ (a) and Z$=0.008$ (b), respectively. Solid lines represent the magnitude limit ($R=21.0$) of this study.
\label{mass}}
\end{figure}

Figure \,\ref{mass} shows the masses versus the ages of the star clusters. 
The masses of the star clusters in IC 10 are ranging from $\sim 10^{2.5} M_{\odot}$ to $\sim 10^{6.0} M_{\odot}$. 
The lower limit of the masses of the star clusters is due to the survey limit of this study.
Young star clusters ($\leq10$ Myr) are mostly less massive than $10^4 M_{\odot}$, while the old star clusters ($>1$ Gyr) are as massive as $\sim 10^5 M_{\odot}$.
The masses of the old star clusters are similar to those of typical globular clusters in the Milky Way, and the mass range of the young star clusters is  similar to that of the young open clusters in the Milky Way.
It is noted that young star clusters have low masses (M$\lesssim10^4$M$_\odot$), so stochastic effects can affect the age estimation of these clusters \citep{Mai09,And13}. This stochastic effect increases errors of age estimation for the range between 10 Myr and 100 Myr \citep{And13}. Therefore, we should be careful to study the star clusters with this range.

\begin{figure}
 \epsscale{1.0} \plotone{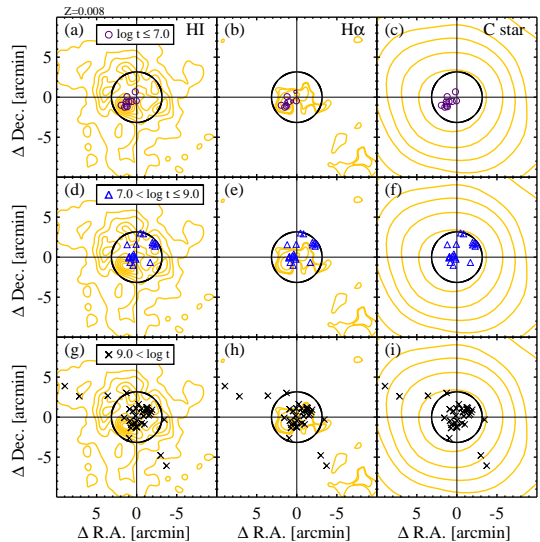}
 \caption{Spatial distribution of the star clusters with different ages in IC 10 derived for Z$=0.008$. Contours in left, middle, and right columns represent the HI map \citep{van02}, H$\alpha$ map \citep{Mas07}, and number density contour map of C stars \citep{Dem04}, respectively.
\label{spaage2}}
\end{figure}

Figure \ref{spaage2} displays spatial distributions of the star clusters with different ages (young, $\leq10$ Myr; intermediate-age, 10 Myr -- 1 Gyr; and old, $>1$ Gyr).
We compared these spatial distributions with the HI map \citep{van02}, H$\alpha$ map \citep{Mas07}, and C star density map \citep{Dem04}
(contours in the figure). 
The H$\alpha$ map represents recent star-forming regions, while the map of C star indicates the spatial distribution of relatively old stars ($\sim1$ Gyr).

The peak position of the young star clusters ($\leq10$ Myr) is consistent with that of the H$\alpha$ emission, which is close to the HI peak position.
This indicates that the recent starburst took place in this region.
The intermediate-age star clusters (10 Myr -- 1 Gyr) are distributed in two groups around the main body of IC 10. One is close to, but about one arcmin from the peak position of the H$\alpha$ emission. The other is about four arcmin from the peak position of the H$\alpha$ emission.
The old star clusters ($>1$ Gyr) are not only distributed in the main body of IC 10 but also in the halo region. 
The old star clusters in the main body shows relatively loose concentration compared with the young star clusters, and their peak position is roughly
consistent with the galaxy center.
The spatial distribution of the old star clusters in the halo region is very asymmetric, while that of the C star is approximately circular.
It is noted the the spatial distribution of the C stars is more extended ($15\arcmin$) than that of the star clusters ($10\arcmin$). 
It is expected that these may be more star clusters outside our survey area.

\section{Discussion}

\subsection{Super Star Clusters in IC 10?}

\citet{Hun01} detected 13 star clusters and OB associations from the HST/WFPC2 images covering about a half of the main body of IC 10, finding that the brightest of which (ID 4-1 in her study, ID 54 in this study) has an absolute magnitude of $M_V \approx-10.0$. \citet{Hun01} pointed out that ID 54 is an OB association based on its low stellar number density. From this she pointed out that there are no Super Star Cluster (SSC) in her field. Although we increased the sample size of the star clusters in IC 10, covering a much larger field than that used in \citet{Hun01}, the cluster ID 54 remains to be the brightest in our sample with $M_V \approx -10.6$, which is only 0.6 mag brighter than the value given by \citet{Hun01}. However, this star cluster ID 54 may not be an OB association based on new stellar number density.

\citet{Hun01} derived the stellar number density of this star cluster from the number of resolved stars within the radius of $3\arcsec.09$ in the WF chips of HST/WFPC2 (see her Figure 14) is comparable with those of OB associations in the Milky Way and the Large Magellanic Cloud. However, this value for the stellar number density should be increased for the following reasons. First, the spatial resolution of WF chips \citet{Hun01} used is limited to deblend stars in the star clusters. We checked the F606W and F814W images obtained with HST/ACS WFC (ID: 9683, PI:F. Bauer) that has twice higher spatial resolution than WF chip in WFPC2 to test the effect of the spatial resolution. We derived photometry of resolved stars on the HST/ACS images using ALLSTAR/DAOPHOT packages in IRAF. Photometric zeropoints for each band is obtained from HST/ACS website\footnote{http://www.stsci.edu/hst/acs/analysis/zeropoints/zpt.py}, and the HST/ACS filter systems are transformed to the Johnson system using the relation in \citet{Sir05}. We found 23 stars with $M_V \leq -4$ in the cluster ID 54 using the distance and reddening as those in \citet{Hun01}, and 26 stars with  $M_V \leq -4$ if we adopt the distance and reddening as adopted in this study. With this value  we calculate the stellar number density, 0.07 stars per square parsec. This number of the resolved stars in the cluster is about twice larger than that from \citet{Hun01}, increasing the value of the star number density by a factor of two. This shows that the star number density of cluster ID 54 is much larger than those of typical OB associations in other galaxies (see Figure  14 in \citet{Hun01}), indicating that this cluster is not an OB association but an SSC.

\begin{figure}
 \epsscale{1.0} \plotone{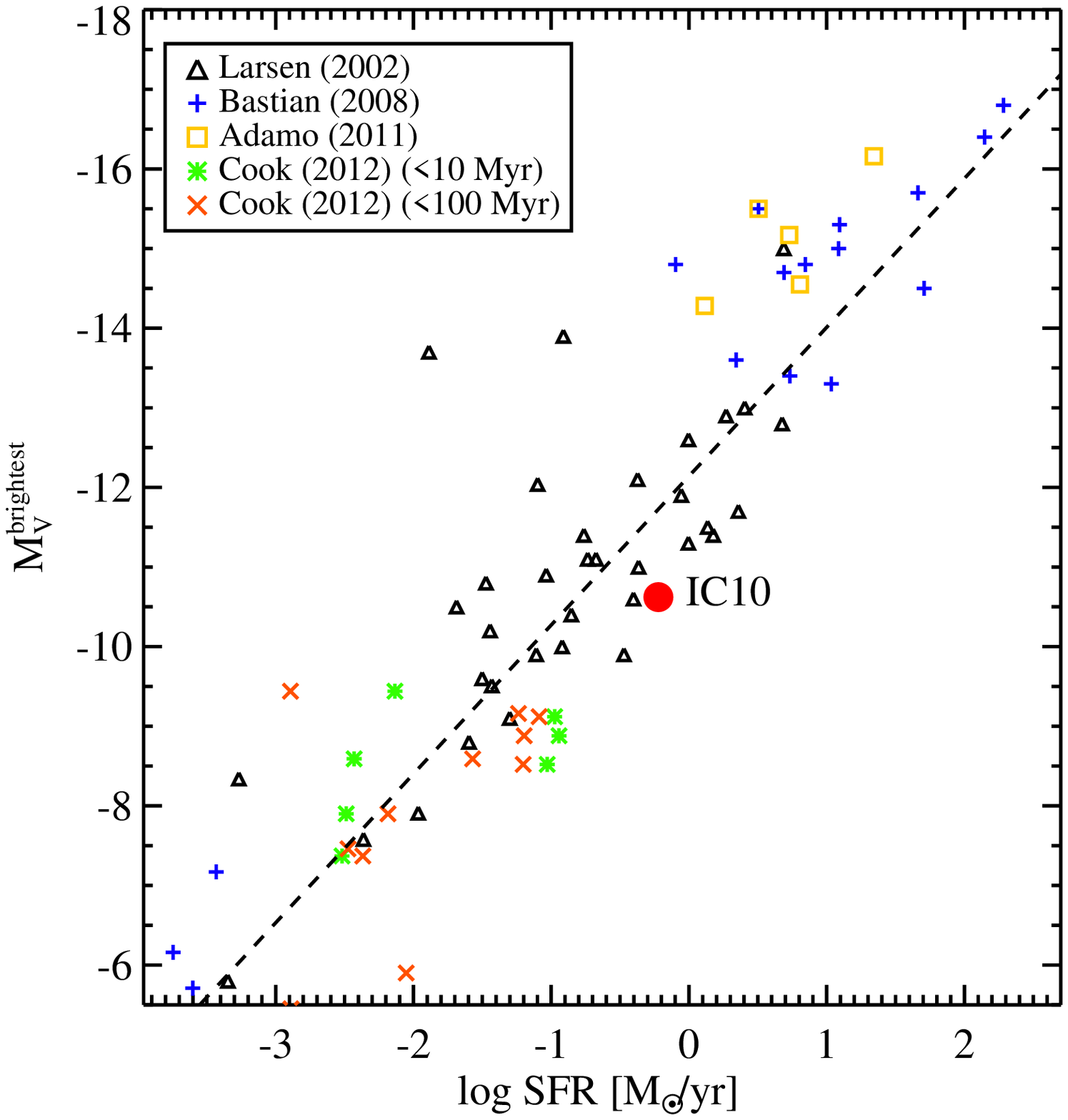}
 \caption{Log star formation rate versus V-band absolute magnitude of the brightest star cluster for nearby galaxies from previous studies \citep{Lar02,Bas08,Ada11,Coo12}. A dashed line shows the fitting line from \citet{Wei04}. The filled circle represents IC 10. The internal and foreground extinctions are corrected for the all V-band absolute magnitudes of the brightest star clusters including that of IC 10 in this figure.
\label{sfrmv}}
\end{figure}

We plotted this SSC on the relation between $V-$band absolute magnitude of the brightest cluster and star formation rate of the host galaxy in Figure \ref{sfrmv} including the results from previous studies \citep{Lar02,Bas08,Ada11,Coo12}. The dashed line in the figure is a linear fit of the samples from \citet{Lar02} given by \citet{Wei04}. If we include ID 54 in our sample as a star cluster, then IC 10 follows well this relation. This result suggests that the masses of the most massive star clusters in galaxies are correlated to the star formation rate of their host galaxies. However, \citet{Ada11} pointed out that the brightest star clusters in BCDs are located about 1.0 magnitude above this relation due to their high cluster formation efficiency (CFE, defined as cluster formation rate/star formation rate by \citealp{Bas08}). IC 10 is known as a BCD, but it follows well the relation between the brightest star clusters and the star formation rate of the host galaxies. Therefore, we need to check the CFE of IC 10.

\begin{figure}
 \epsscale{1.0} \plotone{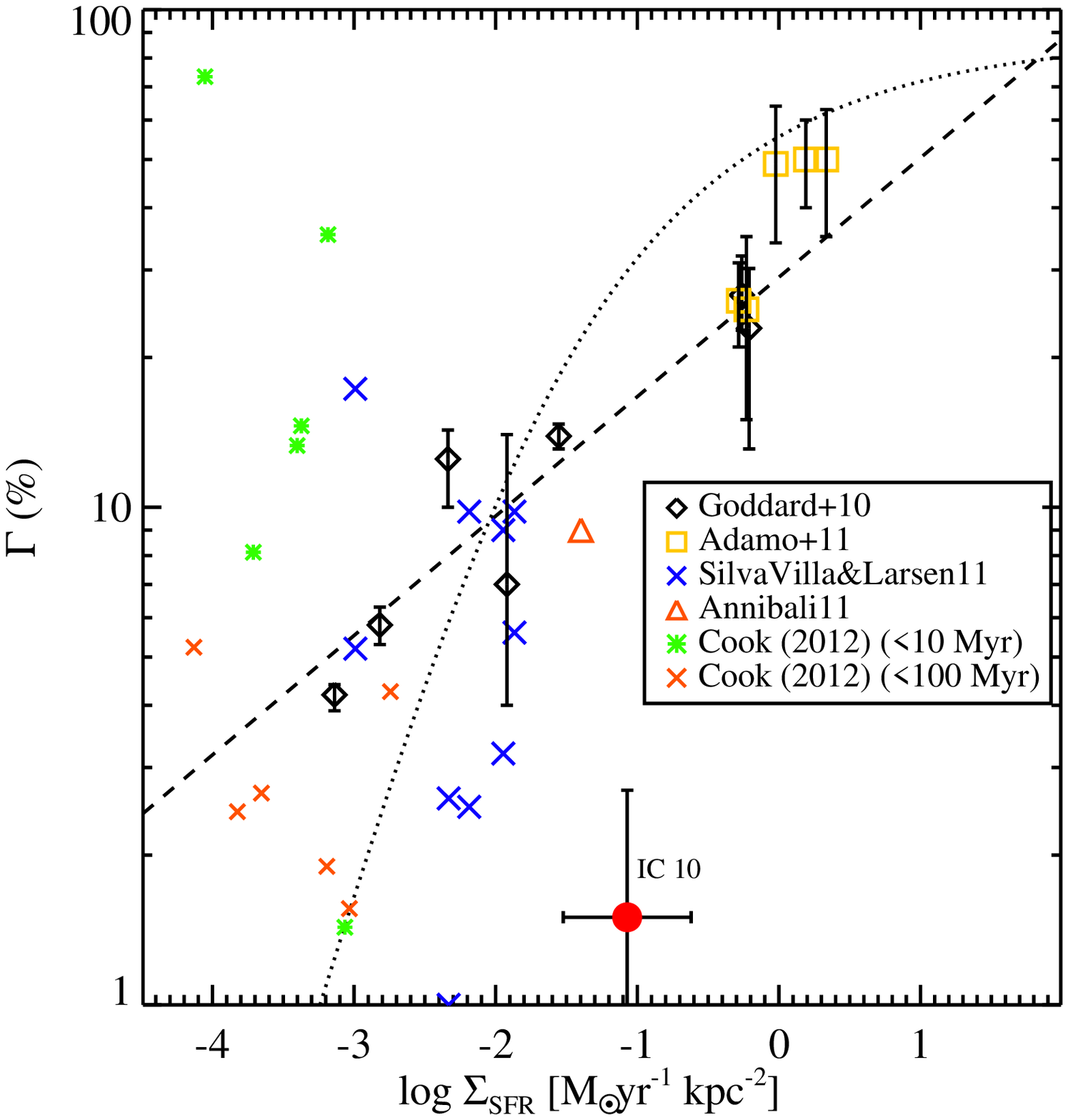}
 \caption{Star cluster formation efficiency versus log surface star formation rate. The filled circle displays IC 10.  The other samples from previous studies \citep{God10,Ada11,Sil11,Ann11,Coo12}. A dashed line shows the fitting result of \citet{God10}. The theoretical expectation of \citet{Kru12} is indicated with a dotted line.
\label{ssfr}}
\end{figure}

Figure \ref{ssfr} shows CFEs versus surface star formation rates of several galaxies from previous studies \citep{God10,Ada11,Sil11,Ann11,Coo12}. CFE is defined as cluster formation rate/star formation rate, where the cluster formation rate is the total star cluster mass divided by the time interval of the experiment (10 Myr in this study). The total cluster mass is estimated as follows. First, we summed star cluster masses that are more massive than $10^3 M_{\odot}$. Second, we assumed the cluster mass function as a power-law function, $N(m)dm \varpropto m^{\beta}$ with $\beta=-2$, and integrated this cluster mass function from $10^3 M_{\odot}$ to $10^2 M_{\odot}$ to derive a value of the total mass for the low mass clusters. Finally, we summed these two values to get a value of the total cluster mass, which was used in this study. Dwarf galaxy samples from \citet{Coo12} are diverse in the relation, and IC 10 also increases the dispersion of this relation. It may be caused by the large error of CFE due to a small number of star clusters. \citet{Coo12} found that the scatters mostly come from the stochastic effect, but several galaxies show significant deviations exceeding the stochastic scatter. The result of IC 10 increases the scatter in the relation. It is noted that IC 10 has the highest surface star formation rate among the outliers in the relation between CFE and surface star formation rate. However, it is hard to constrain a reason for this result of IC 10 due to the limit of our data. Further studies are needed to understand why IC 10 has lower CFE compared with that of other star forming galaxies.

\subsection{The Halo of IC 10}

IC 10 has an extended structure outside the main body, including HI clouds, old stars (C stars and RGB stars), planetary nebulae, and star clusters (\citealp{Wil98,Dem04,San10,Gon12}, and this study).
This extended structure is order of magnitude larger ($r\approx30\arcmin$) than the optical main body ($R_{25} = 3\arcmin.15$).
The extended structure has mostly old stellar populations including old stars and star clusters, while  the main body is dominated by young stellar populations.
This indicates that IC 10 has an old halo, embedding a disk of the main body.
This kind of old stellar halo 
has been found in several other dwarf irregular galaxies (e.g. NGC 6822 \citep{Lee05, Hwa11}, Large Magellanic Cloud \citep{Min03}, Leo A \citep{Van04}, NGC 1569, and NGC 4449 \citep{Rys11}). 

The formation of stellar halos in large and massive galaxies including the Milky Way has been explained in terms of hierarchical merging and accretion \citep{Whi78,Sea78}. For dwarf galaxies, two scenarios were suggested to explain the formation of halos: merging between gas-rich dwarf irregular galaxies \citep{Bek08a} or early star formation activity with supernova feedback \citep{Sti09}. These two scenarios can explain reasonably the existence of extended old populations and HI clouds, but which of the two is more effective is not yet clear.

The linear alignment of the halo star clusters in IC 10 may provide a hint for understanding the halo formation. It is well known that halo globular clusters in giant elliptical galaxies are distributed with a spheroidal shape (e.g \citealp{Lee03}). In addition, the Large Magellanic Cloud that is a dwarf galaxy similar to IC 10 does not show any sub structures of halo globular clusters (e.g. \citealp{Bic08}). Therefore, the spheroidal structure has been regarded as a normal spatial distribution of globular clusters in the halo. 

However, the linear structures or substructures of the halo star clusters are found in several other galaxies. NGC 6822, a dwarf irregular galaxy similar to IC 10 has a linear structure of halo globular clusters \citep{Hwa11}. It is suggested that the halo globular clusters in NGC 6822 may have been accreted in the halo \citep{Hwa14}. The anisotropic spatial distributions of satellite systems were also found in the Milky Way \citep{Hol69,Zar97} and M31 \citep{Iba13}. It has been interpreted in terms of the accretion of satellite galaxies along filaments (e.g. \citealp{Kne04}). A recent study of halo globular clusters in M31 show that the globular cluster system is aligned along the tidal streams, suggesting that the globular clusters may be remnants of recently accreted dwarf galaxies into M31 \citep{Vel14}. Several studies suggested that this kind of accretion can happen in dwarf galaxies (e.g. NGC 4449 \citep{Str12,Mar12}, NGC 6822 \citep{Hwa14}). If we adopt this interpretation to IC 10, the halo star clusters in IC 10 might be metal-poor globular clusters accreted along filaments.

\subsection{The Origin of the Starbursts }

The starburst phenomenon in BCDs such as IC 10 has been a mystery because it is not easy to understand how low mass galaxies such as BCDs can keep gas long and show 
strong starbursts later.
Several processes were proposed for explaining the origin of the starbursts in BCDs: in-spiraling gas clump \citep{Elm12}, gas cloud collision \citep{Gor81}, mergers \citep{Ost01,Bek08a,Bek08b}, galactic winds \citep{Ost03}, and tidal interactions \citep{van98}.

We will discuss the origin of the starbursts in IC 10 using the observational results
found in this study and previous studies.
To explain complex structure and kinematics of HI gas in IC 10,
\citet{Wil98} suggested that the recent starburst started a few $10^7$ yr ago, and is still going on, as the gas keeps infalling from the reservoir in the outer region.

Recently \citet{Nid13} found a new interesting structure from deep HI observation, 
a long (5.2 kpc) HI extension in the north-west region.
They pointed out that this HI extension is not related with the recent starburst,
but may be with an interaction with another dwarf galaxy before this starburst.
However, they did not mention the epoch of the dwarf galaxy interaction.
The velocity of the HI extension with respect to IC 10 is about 65 km s$^{-1}$. 
Assuming that the HI extension move with this velocity in the projected plane, we estimate that it would take roughly 300 Myr for this to move from IC 10 to the current position (18 kpc). 

We found from the age distribution of the clusters that there were two starbursts,
one at $6$ Myr and another at $100$ Myr.
The first burst at $100$ Myr produced star clusters mostly in two regions, one in the central star-forming region and another in the north-west star-forming region. 
The second burst at 6 Myr formed star clusters mostly in the south-east star-forming region.

It is noted that the separation of these two age peaks can be caused by the limit of SED fitting method. Theoretical SSP models have a degeneracy around log$(age)=7$, so the SED fitting method does not recover well around the age of log$(age)=7$ (e.g. \citealp{Mai09}). In addition, many star clusters in IC 10 are less massive than $10^4$ M$_{\odot}$ , so they also have the stochastic effect of IMF sampling. Because of the above reason, there is a possibility that these two peaks may be originally from one broad distribution. However, there is still a possibility that these two peaks can be genuine separated peaks. The spatial distributions of these two peaks are different, and these peaks are shown in both age distributions with normal histogram and dN/dt. The color-color diagram of star clusters in IC 10 also suggests the presence of a gap in star cluster formation around 10 Myr (Figure \ref{ubvccd}b). In addition, the ages of these peaks are consistent with the epochs of recent starburst and past interaction suggested by previous studies. Therefore, we regard these two age peaks as probable separated peaks.

Star cluster disruption can also affect the age distribution of star clusters. There are two models for the cluster disruption that are mass dependent disruption (MDD, \citealp{Lam05}) and mass independent disruption (MID, \citealp{Whi07}) models. It is hard to constrain with our sample which cluster disruption model works better for IC10, and how much star clusters were disrupted in IC10. However, both disruption models may not affect our interpretation for the age distribution because of the following reasons. If the time scale of cluster disruption is long ($\geq 10^8$yr), then the cluster disruption has been going gradually. Therefore, the age distribution may not change drastically. If the time scale of cluster disruption is short ($\leq 10^7$ yr), then star clusters older than 10 Myr remain to be a small fraction of them (e.g. \citealp{Cha10}). In this case, the starburst at 100 Myr may be stronger than the starburst at 6 Myr, but the age peaks may not disappear. Therefore, cluster disruption may not remove the peaks in the age distribution.

The studies of HI gas in IC 10 and this study have links to understand the recent starbursts. 
The first starburst epoch found in this study corresponds to the encounter epoch of the HI extension estimated with its radial velocity and projected distance to IC 10. 
In addition, the location of the star clusters belonging to the first starburst is closer to the HI extension than those of other star clusters. 
Therefore, it is concluded that the HI extension was formed when IC 10 interacted with another gas-rich dwarf galaxy $\sim$100 Myr ago.

It is not clear how the first starburst propagates to the second starburst in this study. 
\citet{Wil98} suggested that the stellar wind from massive stars formed at the first starburst in this study triggers the second starburst by compressing the gas. 
However, the time scale that they mentioned is a little different with the time interval between the first and second starbursts in this study. 
\citet{Bek08b} mentioned that the dwarf-dwarf interaction can make a BCD and there may be two starbursts in this case, though he did not explain the mechanism for two starbursts.

From above, we suggest a  scenario as follows.
Two gas-rich dwarf galaxies tidally interacted at $\sim$100 Myr ago, 
and the remnant of this encounter traveled to the northwest, leaving a long HI extension  \citep{Bek08b,Nid13}.
At this time, a starburst occurred in the north-west star forming region, forming several star clusters. 
Massive stars in these star clusters affected the interstellar medium through stellar winds or supernova explosion. 
Then the second starburst happened at 6 Myr ago, and it is the recent starburst of IC 10.

\section{Summary}

We presented a photometric study of 66 star clusters in IC 10 including five new star clusters found in this study. 
The integrated $UBVRI$ magnitudes of these star clusters were obtained, and ages of them were estimated by the SED fitting method. 
Our primary results are summarized below.
\begin{enumerate}
\item The number density profiles of the star clusters show a break at $R_{25}$ and we divided the star clusters in IC 10 into two groups: main body clusters at $R<R_{25}$ and halo star clusters at $R\geq R_{25}$.
All five star clusters found in this study are halo star clusters.
The halo star clusters in IC 10 shows a linear structure, and this structure seems to be correlated with the extended HI disk. 
The halo star clusters are mostly red ($(B-V)>1.2$) and old ($>1$ Gyr), while the disk star clusters have various color ranges ($0.5\lesssim (B-V) \lesssim 2.0$) and age ranges (5 Myr -- 10 Gyr). 

\item The age distribution of the star clusters in IC 10 shows three peaks at $\sim$6 Myr, $\sim$100 Myr, and $\sim$4 Gyr. The youngest peak is consistent with the recent starburst, and the intermediate-age peak suggests the existence of another starburst.

\item The spatial distribution of the young star clusters ($<10$ Myr) are well consistent with H$\alpha$ emission and concentrated on the small region in the main body, while the old star clusters ($>1$ Gyr) are distributed in a wider area than the disk.
These young star clusters can be a result of the recent starburst event.
It seems that the spatial distribution of the intermediate-age star clusters is not correlated with the structures of HI gas, H$\alpha$ emission, and C stars. 
However, they seem to be related with past merger or tidal interaction with HI cloud at the northwest.

\item One SSC is found in IC 10. The brightest star cluster (ID 54) in IC 10 is $M_V\approx-10.6$ after correction of foreground and internal reddenings. 
It has relatively low stellar number density compared with those of typical SSCs.

\item The $(B-V)$ color-histograms and the ages of the halo star clusters suggests that most halo star clusters may be metal-poor old star clusters, and it similar to that of halo star clusters in other starburst galaxies (e.g. NGC 4449 and M82) and normal spiral galaxies (e.g. M31 and Milky Way). 
The anisotropic spatial distribution of the halo star clusters and the existence of these old metal-poor halo star clusters suggest that the halo of IC 10 may be formed by accretions.

\item We found a group of intermediate-age star clusters with $\sim 100$ Myr, and it may be related with the HI extension at the northwest. 
This HI extension might have interacted with IC 10 at $\sim$100 Myr ago, suggesting that the tidal interaction between two gas-rich dwarf galaxies occurred at $\sim$100 Myr ago. 
This tidal interaction led to the starburst in IC 10 at $\sim$100 Myr ago, and the stellar feed back process might have caused the recent starburst in IC 10.

\end{enumerate}

\acknowledgments 

We thank the anonymous referee for helpful comments which improved the original manuscript.
This work was supported by the National Research Foundation of Korea (NRF) grant
funded by the Korea Government (MSIP) (No.2013R1A2A2A05005120).




\makeatletter
\clearpage
\begin{table}
\tiny
\begin{center}
\caption{A catalog of the star clusters in IC 10$^{\rm a}$ \label{tbl-2}}
\begin{tabular}{lllccccccccccccc}
\tableline\tableline ID & R.A. & Dec. & $V$ & ($B-V$) &
 ($V-R$) & ($R-I$) & ($U-B$) & log(age) & E(B-V) (Z$=0.008$) & log mass & M$_V$ & $r_{\rm eff}$ \\
   & (J2000) & (J2000) & [mag] &  &  & &  & [yr] & & M$_{\odot}$ & [mag] & [pc]\\
\tableline
%
%
%
%

  1 &   4.97476238 &  59.22408231 & 18.99 $\pm$  0.01 &  1.30 $\pm$  0.02 &  0.85 $\pm$  0.01 &  0.87 $\pm$  0.01 &  0.49 $\pm$  0.03 &  9.69$^{+ 0.02}_{- 0.09}$ &  0.32$^{+ 0.01}_{- 0.01}$ &  5.18 &  -7.88 &  3.68 \\
  2 &   4.99087157 &  59.33122718 & 20.57 $\pm$  0.04 &  0.99 $\pm$  0.07 &  0.59 $\pm$  0.06 &  0.85 $\pm$  0.06 &  0.20 $\pm$  0.08 &  8.09$^{+ 0.08}_{- 0.04}$ &  0.60$^{+ 0.00}_{- 0.03}$ &  3.63 &  -7.17 &  3.16 \\
  3 &   4.99725978 &  59.32664959 & 21.33 $\pm$  0.05 &  0.90 $\pm$  0.08 &  0.74 $\pm$  0.07 &  1.19 $\pm$  0.06 &  0.02 $\pm$  0.08 &  8.11$^{+ 0.40}_{- 0.04}$ &  0.60$^{+ 0.00}_{- 0.04}$ &  3.41 &  -6.42 & ... \\
  4 &   5.00011749 &  59.33282947 & 19.85 $\pm$  0.02 &  1.16 $\pm$  0.03 &  0.82 $\pm$  0.02 &  0.90 $\pm$  0.02 &  0.64 $\pm$  0.05 &  9.00$^{+ 0.02}_{- 0.02}$ &  0.54$^{+ 0.01}_{- 0.00}$ &  4.48 &  -7.71 & ... \\
  5 &   5.00802291 &  59.32931640 & 20.27 $\pm$  0.03 &  1.00 $\pm$  0.04 &  0.69 $\pm$  0.03 &  0.83 $\pm$  0.03 &  0.29 $\pm$  0.05 &  8.34$^{+ 0.04}_{- 0.13}$ &  0.60$^{+ 0.00}_{- 0.02}$ &  3.92 &  -7.48 &  4.41 \\

\tableline
\tablenotetext{1}{(This table is available in its entirety in a machine-readable form in the online journal. A portion is shown here for guidance regarding its form and content.)}
\end{tabular}
\end{center}
\end{table}


\begin{thebibliography}{}
\bibitem[Adamo et al.(2011)]{Ada11} Adamo, A., {\"O}stlin, 
G., \& Zackrisson, E.\ 2011, \mnras, 417, 1904 


\bibitem[Anders et al.(2013)]{And13} Anders, P., Kotulla, R., 
de Grijs, R., \& Wicker, J.\ 2013, \apj, 778, 138


\bibitem[Annibali et al.(2011)]{Ann11} Annibali, F., Tosi, 
M., Aloisi, A., \& van der Marel, R.~P.\ 2011, \aj, 142, 129 




\bibitem[Bastian(2008)]{Bas08} Bastian, N.\ 2008, \mnras, 
390, 759 


\bibitem[Bekki(2008a)]{Bek08a} Bekki, K.\ 2008, \apjl, 680, L29 


\bibitem[Bekki(2008b)]{Bek08b} Bekki, K.\ 2008, \mnras, 388, 
L10 


\bibitem[Bertin 
\& Arnouts(1996)]{Ber96} Bertin, E., \& Arnouts, S.\ 1996, \aaps, 117, 393 


\bibitem[Bica et al.(2008)]{Bic08} Bica, E., Bonatto, C., 
Dutra, C.~M., \& Santos, J.~F.~C.\ 2008, \mnras, 389, 678

\bibitem[Billett et al.(2002)]{Bil02} Billett, O.~H., Hunter, 
D.~A., \& Elmegreen, B.~G.\ 2002, \aj, 123, 1454 


\bibitem[Borissova et 
al.(2000)]{Bor00} Borissova, J., Georgiev, L., Rosado, M., et al.\ 2000, \aap, 363, 130 


\bibitem[Bruzual 
\& Charlot(2003)]{Bru03} Bruzual, G., \& Charlot, S.\ 2003, \mnras, 344, 1000 


\bibitem[Chandar et al.(2010)]{Cha10} Chandar, R., Fall, 
S.~M., \& Whitmore, B.~C.\ 2010, \apj, 711, 1263 


%
%




\bibitem[Cook et al.(2012)]{Coo12} Cook, D.~O., Seth, A.~C., 
Dale, D.~A., et al.\ 2012, \apj, 751, 100 



\bibitem[Crone et al.(2002)]{Cro02} Crone, M.~M., 
Schulte-Ladbeck, R.~E., Greggio, L., \& Hopp, U.\ 2002, \apj, 567, 258 



\bibitem[de Vaucouleurs et al.(1991)]{deV91} de Vaucouleurs,
G., de Vaucouleurs, A., Corwin, H.~G., Jr., et al.\ 1991, Third Reference Catalogue of Bright Galaxies, Vols. 1-3 (New York: Springer), 2069

\bibitem[Demers et 
al.(2004)]{Dem04} Demers, S., Battinelli, P., \& Letarte, B.\ 2004, \aap, 424, 125 


\bibitem[Drozdovsky et al.(2001)]{Dro01} Drozdovsky, I.~O., 
Schulte-Ladbeck, R.~E., Hopp, U., Crone, M.~M., 
\& Greggio, L.\ 2001, \apjl, 551, L135 


\bibitem[Elbaz 
\& Cesarsky(2003)]{Elb03} Elbaz, D., \& Cesarsky, C.~J.\ 2003, Science, 300, 270 


\bibitem[Elmegreen et al.(2012)]{Elm12} Elmegreen, B.~G., 
Zhang, H.-X., \& Hunter, D.~A.\ 2012, \apj, 747, 105 


\bibitem[Garnett(1990)]{Gar90} Garnett, D.~R.\ 1990, \apj, 
363, 142 

\bibitem[Georgiev et 
al.(1996)]{Geo96} Georgiev, T.~B., Tikhonov, N.~A., \& Karachentsev, I.~D.\ 1996, Astronomical and Astrophysical Transactions, 11, 47 


\bibitem[Gil de Paz et al.(2003)]{Gil03} Gil de Paz, A., 
Madore, B.~F., \& Pevunova, O.\ 2003, \apjs, 147, 29 


\bibitem[Goddard et al.(2010)]{God10} Goddard, Q.~E., 
Bastian, N., \& Kennicutt, R.~C.\ 2010, \mnras, 405, 857 



\bibitem[Gon{\c c}alves et al.(2012)]{Gon12} Gon{\c c}alves, 
D.~R., Teodorescu, A.~M., Alves-Brito, A., M{\'e}ndez, R.~H., 
\& Magrini, L.\ 2012, \mnras, 425, 2557 


\bibitem[Gordon 
\& Gottesman(1981)]{Gor81} Gordon, D., \& Gottesman, S.~T.\ 1981, \aj, 86, 161 







\bibitem[Hodge \& Lee (1990)]{Hod90}Hodge, P. W., \& Lee, M. G. 1990, \pasp, 102, 26

\bibitem[Holmberg(1969)]{Hol69} Holmberg, E.\ 1969, Arkiv for 
Astronomi, 5, 305 




\bibitem[Hunter(2001)]{Hun01} Hunter, D.~A.\ 2001, \apj, 559, 
225 

\bibitem[Hwang \& Lee(2008)]{Hwa08} Hwang, N., \& Lee, M.~G.\ 2008, \aj, 135, 1567 


\bibitem[Hwang et al.(2011)]{Hwa11} Hwang, N., Lee, M.~G., 
Lee, J.~C., et al.\ 2011, \apj, 738, 58 


\bibitem[Hwang et al.(2014)]{Hwa14} Hwang, N., Park, H.~S., 
Lee, M.~G., et al.\ 2014, \apj, 783, 49 


\bibitem[Ibata et al.(2013)]{Iba13} Ibata, R.~A., Lewis, 
G.~F., Conn, A.~R., et al.\ 2013, \nat, 493, 62 



\bibitem[Karachentsev 
\& Tikhonov(1993)]{Kar93} Karachentsev, I.~D., \& Tikhonov, N.~A.\ 1993, \aaps, 100, 227 


\bibitem[Kim et al.(2009)]{Kim09} Kim, M., Kim, E., Hwang, 
N., et al.\ 2009, \apj, 703, 816 

\bibitem[Knebe et al.(2004)]{Kne04} Knebe, A., Gill, 
S.~P.~D., Gibson, B.~K., et al.\ 2004, \apj, 603, 7 






\bibitem[Kruijssen(2012)]{Kru12} Kruijssen, J.~M.~D.\ 2012, 
\mnras, 426, 3008 


\bibitem[Lamers et 
al.(2005)]{Lam05} Lamers, H.~J.~G.~L.~M., Gieles, M., Bastian, N., et al.\ 2005, \aap, 441, 117 



\bibitem[Larsen(1999)]{Lar99} Larsen, S.~S.\ 1999, \aaps, 139, 393 


\bibitem[Larsen(2002)]{Lar02} Larsen, S.~S.\ 2002, \aj, 124, 
1393 




\bibitem[Lee(2003)]{Lee03} Lee, M.~G.\ 2003, Journal of 
Korean Astronomical Society, 36, 189 


\bibitem[Lee 
\& Hwang(2005)]{Lee05} Lee, M.~G., \& Hwang, N.\ 2005, IAU Colloq.~198: Near-fields cosmology with dwarf elliptical galaxies, 181 



\bibitem[Leitherer et al.(1999)]{Lei99} Leitherer, C., 
Schaerer, D., Goldader, J.~D., et al.\ 1999, \apjs, 123, 3 


\bibitem[Lim et al.(2013)]{Lim13} Lim, S., Hwang, N., 
\& Lee, M.~G.\ 2013, \apj, 766, 20 


\bibitem[Magrini et 
al.(2003)]{Mag03} Magrini, L., Corradi, R.~L.~M., Greimel, R., et al.\ 2003, \aap, 407, 51 




\bibitem[Ma{\'{\i}}z Apell{\'a}niz(2009)]{Mai09} Ma{\'{\i}}z 
Apell{\'a}niz, J.\ 2009, \apj, 699, 1938

\bibitem[Mart{\'{\i}}nez-Delgado et al.(2012)]{Mar12} 
Mart{\'{\i}}nez-Delgado, D., Romanowsky, A.~J., Gabany, R.~J., et al.\ 
2012, \apjl, 748, L24 



\bibitem[Massey 
\& Armandroff(1995)]{Mas95} Massey, P., \& Armandroff, T.~E.\ 1995, \aj, 109, 2470 

\bibitem[Massey \& Holmes (2002)]{Mas02} Massey, P., \& Holmes, S. 2002, \apj, 580, L35 

\bibitem[Massey et al.(2007)]{Mas07} Massey, P., Olsen, 
K.~A.~G., Hodge, P.~W., et al.\ 2007, \aj, 133, 2393 


\bibitem[McQuinn et al.(2010)]{McQ10} McQuinn, K.~B.~W., 
Skillman, E.~D., Cannon, J.~M., et al.\ 2010, \apj, 721, 297 


\bibitem[Melisse 
\& Israel(1994)]{Mel94} Melisse, J.~P.~M., \& Israel, F.~P.\ 1994, \aap, 285, 51 


\bibitem[Meurer et al.(1995)]{Meu95} Meurer, G.~R., Heckman, 
T.~M., Leitherer, C., et al.\ 1995, \aj, 110, 2665 


\bibitem[Minniti et al.(2003)]{Min03} Minniti, D., Borissova, 
J., Rejkuba, M., et al.\ 2003, Science, 301, 1508 





\bibitem[Nidever et al.(2013)]{Nid13} Nidever, D.~L., Ashley, 
T., Slater, C.~T., et al.\ 2013, \apjl, 779, L15 




\bibitem[{\"O}stlin et 
al.(2001)]{Ost01} {\"O}stlin, G., Amram, P., Bergvall, N., et al.\ 2001, \aap, 374, 800 


\bibitem[{\"O}stlin et 
al.(2003)]{Ost03} {\"O}stlin, G., Zackrisson, E., Bergvall, N.,  R{\"o}nnback, J.\ 2003, \aap, 408, 887 








\bibitem[Richer et 
al.(2001)]{Ric01} Richer, M.~G., Bullejos, A., Borissova, J., et al.\ 2001, \aap, 370, 34 


\bibitem[Ry{\'s} et 
al.(2011)]{Rys11} Ry{\'s}, A., Grocholski, A.~J., van der Marel, R.~P., Aloisi, A., \& Annibali, F.\ 2011, \aap, 530, A23 
%
%
%
%



\bibitem[Sanna et al.(2010)]{San10} Sanna, N., Bono, G., 
Stetson, P.~B., et al.\ 2010, \apjl, 722, L244 



\bibitem[Searle 
\& Zinn(1978)]{Sea78} Searle, L., \& Zinn, R.\ 1978, \apj, 225, 357 


\bibitem[Sharina et al.(2010)]{Sha10} Sharina, M.~E., 
Chandar, R., Puzia, T.~H., Goudfrooij, P., 
\& Davoust, E.\ 2010, \mnras, 405, 839 


\bibitem[Silva-Villa 
\& Larsen(2011)]{Sil11} Silva-Villa, E., \& Larsen, S.~S.\ 2011, \aap, 529, AA25 





\bibitem[Sirianni et al.(2005)]{Sir05} Sirianni, M., Jee, 
M.~J., Ben{\'{\i}}tez, N., et al.\ 2005, \pasp, 117, 1049 


\bibitem[Skillman et al.(1989)]{Ski89} Skillman, E.~D., 
Kennicutt, R.~C., \& Hodge, P.~W.\ 1989, \apj, 347, 875 


\bibitem[Stinson et al.(2009)]{Sti09} Stinson, G.~S., 
Dalcanton, J.~J., Quinn, T., et al.\ 2009, \mnras, 395, 1455 



\bibitem[Strader et al.(2012)]{Str12} Strader, J., Seth, 
A.~C., \& Caldwell, N.\ 2012, \aj, 143, 52 


\bibitem[Tikhonov 
\& Galazutdinova(2009)]{Tik09} Tikhonov, N.~A., \& Galazutdinova, O.~A.\ 2009, Astronomy Letters, 35, 748 


\bibitem[Vacca et al.(2007)]{Vac07} Vacca, W.~D., Sheehy, 
C.~D., \& Graham, J.~R.\ 2007, \apj, 662, 272 


\bibitem[van der Hulst(2002)]{van02} van der Hulst, J.~M.\ 
2002, Seeing Through the Dust: The Detection of HI and the Exploration of 
the ISM in Galaxies, 276, 84 


\bibitem[Vansevi{\v c}ius et al.(2004)]{Van04} Vansevi{\v 
c}ius, V., Arimoto, N., Hasegawa, T., et al.\ 2004, \apjl, 611, L93 


\bibitem[van Zee et al.(1998)]{van98} van Zee, L., Skillman, 
E.~D., \& Salzer, J.~J.\ 1998, \aj, 116, 1186 


\bibitem[Veljanoski et al.(2014)]{Vel14} Veljanoski, J., 
Mackey, A.~D., Ferguson, A.~M.~N., et al.\ 2014, \mnras, 442, 2929 



\bibitem[Weidner et al.(2004)]{Wei04} Weidner, C., Kroupa, 
P., \& Larsen, S.~S.\ 2004, \mnras, 350, 1503 


\bibitem[White(1978)]{Whi78} White, S.~D.~M.\ 1978, \mnras, 
184, 185




\bibitem[Whitmore et al.(2007)]{Whi07} Whitmore, B.~C., 
Chandar, R., \& Fall, S.~M.\ 2007, \aj, 133, 1067 



\bibitem[Wilcots 
\& Miller(1998)]{Wil98} Wilcots, E.~M., \& Miller, B.~W.\ 1998, \aj, 116, 2363 



%
%

\bibitem[Zaritsky et al.(1997)]{Zar97} Zaritsky, D., Smith, 
R., Frenk, C.~S., \& White, S.~D.~M.\ 1997, \apjl, 478, L53 


\end{thebibliography}
\end{document}